\documentclass[12pt]{article}
\usepackage{cite}
\usepackage[dvipdfm]{graphicx}
\usepackage{amssymb}
\usepackage{amsmath}
\usepackage{epsfig}
\usepackage{dcolumn}
\usepackage{rotating}
\usepackage{color}
\definecolor{rosso}{cmyk}{0,1,1,0.4}
\definecolor{rossos}{cmyk}{0,1,1,0.55}
\definecolor{rossoc}{cmyk}{0,1,1,0.2}
\definecolor{blu}{cmyk}{1,1,0,0.3}
\definecolor{blus}{cmyk}{1,1,0,0.6}
\definecolor{bluc}{cmyk}{1,1,0,0.1}
\definecolor{verde}{cmyk}{0.92,0,0.59,0.25}
\definecolor{verdec}{cmyk}{0.92,0,0.59,0.15}
\definecolor{verdes}{cmyk}{0.92,0,0.59,0.7}

\newcommand{\ba}{\begin{eqnarray}}
\newcommand{\ea}{\end{eqnarray}}
\newcommand{\be}{\begin{equation}}
\newcommand{\ee}{\end{equation}}
\newcommand{\bi}{\begin{itemize}}
\newcommand{\ei}{\end{itemize}}
\newcommand{\la}{\lambda}

\newcommand{\sa}{\sigma}

\newcommand{\Ga}{\Gamma}

\newcommand{\La}{\Lambda}
\newcommand{\ra}{\rightarrow}

\newcommand{\vs}{\vspace{5mm}\\}

\begin{document}

\tolerance=100000
\thispagestyle{empty}
\vspace{1cm}

\begin{center}
{\LARGE \bf
Before the Bang
 \\ [0.15cm]
% \&
% \\ [0.18cm]
% when it is long
}
\vskip 2cm
{\large Tirthabir Biswas}
\vskip 7mm
{ 6300 St. Charles Avenue\\Department of Physics, Box 92\\
Loyola University\\
 New Orleans, LA 70118\\
 {\it tbiswas@loyno.edu} }
\vskip 3mm
\end{center}

\date{\today}

\begin{abstract}
While inflation has been an extremely successful cosmological paradigm,
almost certainly something have had to have happened before it began. Can the
pre-inflationary phase be of any theoretical or phenomenological significance?
Could Quantum Gravity have played any interesting role in this story? These
are some of the questions we want to explore in this essay written for
the Gravity Research Foundation 2013 ``Awards for Essays on Gravitation'' competition. 
\end{abstract}

\newpage

\setcounter{page}{1}
%%%%%%%%%%%%%%%%%%%%%%%%%%%%%%%%%%%%%%%%%%%%%%%%%%%%%
The global dynamics of our observable universe is well approximated by the homogeneous
isotropic FLRW cosmology, the approximation becoming increasing better
as one goes back in the past. This is verified by the "one part in a million" temperature
fluctuations observed in the CMB which formed $\sim$ eV temperatures. According
to General theory of Relativity (GR) and standard fluid/thermo dynamics,
this uniformity should continue as we go further back in time, the success of the Big Bang
Nucleosynthesis (BBN), operating at temperatures of around a MeV, corroborates
this picture.

Now, GR itself has only been tested directly up to around $\mu m\sim (mev)^{-1}$~\cite{short-distance-geraci,short-distance-weld,short-distance-newman}, and therefore gravity may yet surprise us, say, before we hit TeV. However, the phenomenological success of inflation~\cite{WMAP,Planck}  gives us another reason to believe in GR potentially all the way up to  $\sim 10^{14}$ GeV, the typical scale of inflation: the generation of nearly scale-invariant fluctuations  and it's subsequent propagation to the CMB epoch, the phase of inflation itself and the subsequent conversion of inflaton energy into particle excitations through reheating, are all primarily based on GR in the broad sense (which may include extra scalars, for instance). Thus, if we continue to trust GR all the way up to $\sim 10^{14}$ GeV, at this epoch our observable universe had to have been around the size of a soccer ball, and more importantly, still containing only tiny fluctuations in energy
density. For a perfectly efficient ``reheating mechanism'',  $10^{14}$ GeV is also approximately the reheat temperature and in many ways it is this ``moment'' which can be considered as the hot ``Bang''
of modern cosmology; beyond this we supposedly had a cold and dark universe. 

So,
what happened before this new Bang? Well, according to the standard lore, at least around 60 efoldings of  near
exponential inflationary growth which takes our observable universe at least down to the size
of $\sim 10^{-25}$ cm, still mostly homogeneous and isotropic. How far back can we continue to go?

This is where conjectures start to creep in as the threat of the Big Bang singularity looms large. Although the inflationary growth, $a(t)\sim e^{\la t}$, may seem to be able to avoid the singularity by stretching time all the way back to past infinity, this is only an illusion as illustrated in some of the classic papers by Guth, Borde, Vilenkin and Linde~\cite{Guth,Borde,Linde}: inflationary space-times are not geodesically complete in the past, if we
follow any particle trajectory, at some finite proper time in the past the particle trajectory ends abruptly; something else had to have occurred before inflation! In fact, the separation of
scales between the inflationary dynamics and the Planck scale ($\sim 10^{19}$ GeV) suggests a
pre-inflationary phase where the GR description of space-time was still valid. Perhaps,
we just had a matter/radiation/stringy phase lasting all the way to the Planck scale
where the "effective" space-time description of our patch broke down, and we simply
accept the fact that we don't yet know the language of the Planckian era. The Big
Bang singularity is not resolved, the question of how or why time (or more generally
any dynamics) began is left unanswered, but in spite of these pesky questions this is
probably the best cosmological paradigm we have got!

Let us though humor ourselves for the next few pages; what if an effective/approximate spacetime
description is always possible, what if Quantum Gravity (QG) simply provides
us with a ``Bounce'' where a phase of contraction smoothly transitions into a phase
of expansion whenever some Planckian energy density is breached, see~\cite{BMS,Ashtekar,loop,Shtanov,Freese,Baum,palatini-bounce} for efforts
to realize such mechanisms. The "Big Bounce" now becomes the new Bang, or does it?

Let us step back for a second and try to guess what our observational patch
would look like at Planckian energy densities. Note, the FRW metric should still be
a reasonable description of our cosmology because if the anisotropies were too large
we would possibly forever be stuck in a chaotic Mixmaster behavior, while if the
inhomogeneities were too large, ``all'' of us would likely be sitting inside a black hole!
So a priori, what could a Planckian patch contain? (i) Vacuum energy, $\La$, which
could be positive or negative, (ii) Spatial curvature, $\rho_k\sim a^{-2}$, which again could
contribute positively (open) or negatively (closed) to the energy budget, (iii) some
massless degrees of freedom or radiation, $\rho_r\sim a^{-4}$, (iv) perhaps some massive modes
as well, $\rho_m\sim a^{-3}$, let's just stick to these for simplicity.

If $\La>0$, then it is obvious that for an open or a flat universe, we have what
is called a bouncing cosmology where way back in the past our universe starts out
infinite and in a phase of contraction. Now, one of the virtues of inflation was to be able to
wash away any ``initial condition effect'' by expanding the universe exponentially and
making it uniform, but now one has to first arise at a sufficiently smooth universe
at the bounce point starting from a large contracting universe. This requires an
impossible fine-tuning, let us remind our readers that anisotropies increase as $ a^{-6}$ 
and inhomogenieties are not far behind; one of the biggest virtues of inflation seems
to be under a serious threat! Perhaps our universe is closed. Unfortunately, this
does not change the scenario. Either the negative curvature density always remains
small as compared to all the other components, or it cancels all the positive energy
components before the quantum bounce. In either case we are stuck with a bouncing
universe, only in the latter case the bounce is mediated by curvature and not QG.

In the bouncing scenario it  also becomes unclear why one should use the Bunch-Davis vacuum initial conditions for the calculation of  the primordial spectrum of fluctuations seeding the  CMBR anisotropies. These calculations agree remarkably well  with the observations~\cite{WMAP,Planck}, but if one has a prior contracting phase, one ought to start with initial seed fluctuations in that phase, track it through the bounce, and then along the inflationary expansion; there are no general arguments known to the author which demonstrate that the resulting spectrum in such a case will still be nearly scale-invariant as observed in CMBR. In fact, cosmologists have been looking at non-inflationary mechanisms to generate the desired near scale-invariant spectrum utilizing the contraction/bounce phase, some notable examples being the ekpyrotic~\cite{ekpyrotic}, matter-bounce~\cite{matter-bounce} and Hagedorn-bounce~\cite{BBMS} scenarios.
So, have we run out of options?

What if $\La<0$? If $|\La|> \rho_r+\rho_m+\rho_k$, FRW evolution is inconsistent, most likely
the universe would be stuck in a static anti-de Sitter like vacuum with massless and
massive excitations, not our kind of universe. So, let us look at $|\La|< \rho_r+\rho_m+\rho_k$ case. As the universe expands, all the matter components dilute and eventually
cancel the negative cosmological constant causing the universe to "turnaround" and
start contracting. The contraction lasts till the universe bounces back to expansion
at Planckian energy densities, the story repeating itself periodically. The trouble
however is that, ever since the advent of supernova data~\cite{supernova}, having a small negative $\La$ is no longer an option, and a large $|\La|$ only makes matters worse. For instance, if we
take $|\La|\sim (10^{14} GeV)^4$ motivated by string/GUT scale physics, each of these cycles
would only last a very short time, $\tau\sim M_p/\sqrt{\La}\sim 10^{-33}$ s, an impossible cosmology
for us to ever exist in.

Did we miss something? Interactions between different species? That generally
creates entropy which can only increase monotonically thereby breaking the periodicity
of the evolution. This was precisely what Tolman pointed out in the 1930's giving
rise to Tolman's famous entropy problem~\cite{tolman1,tolman2}. Ironically, that now turns out to be
a great savior! As discussed in~\cite{cyclic-inflation}, entropy tends to increase by the same factor in every cycle while the time periods of the cycles remain a constant being governed by
$\La$. Therefore this leads to an overall inflationary growth.

A simple illustrative example is to consider some massive species interacting with
massless particles via scattering and decay processes. In a given cycle, equilibrium
can be maintained up to a certain temperature in the expanding branch, below which
the massive species falls out of equilibrium. A subsequent out-of-equilibrium decay
into radiation creates entropy. After the turnaround, once the temperature becomes
sufficiently high, the massive particles are recreated from radiation via the scattering
processes re-establishing thermal equilibrium before the next cycle commences. The
dynamics of matter density, $\rho_m$, can be captured by Boltzman equation of the form
\be
\dot{\rho}_m+3H\rho_m=-\Ga \rho_m+{\sa(T)\over m}[\bar{\rho}_m^2-\rho_m^2]\ ,
\label{rho_m}
\ee
where $\sa$ is the scattering cross-section, $\Ga$ is the decay rate, and $\bar{\rho}$ is the equilibrium matter density. Fig. 1 (left) shows the numerical behavior of $\rho_m$ in a single cycle~\cite{BD}
giving rise to an exponential growth over the course of many many cycles~\cite{BD}, see
Fig. 1 (right).
%%%%%%%%%%%%%%%%%%%%%%%%%%
\begin{figure}[htbp]
\begin{center}
\includegraphics[width=0.45\textwidth,angle=0]{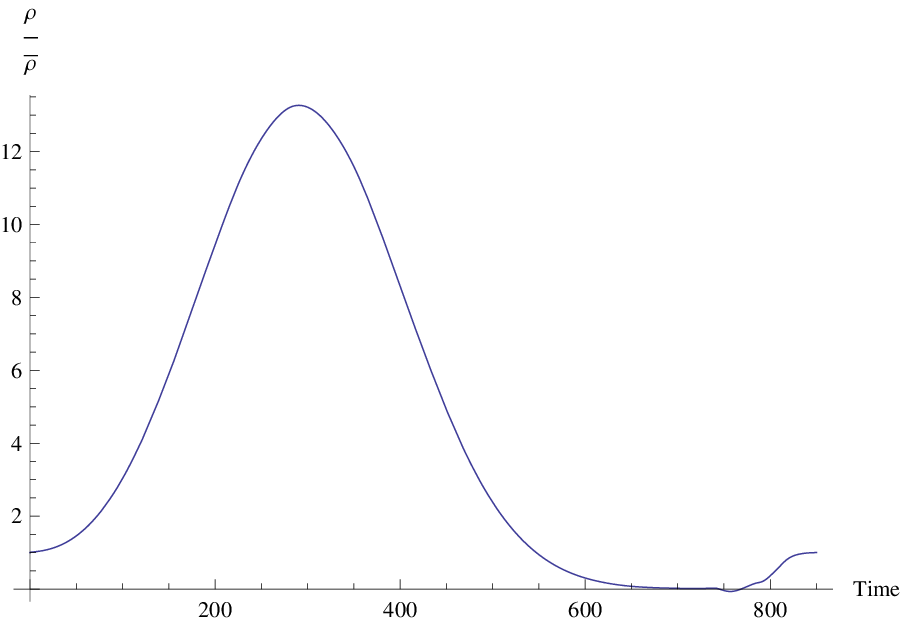}
\includegraphics[width=0.45\textwidth,angle=0]{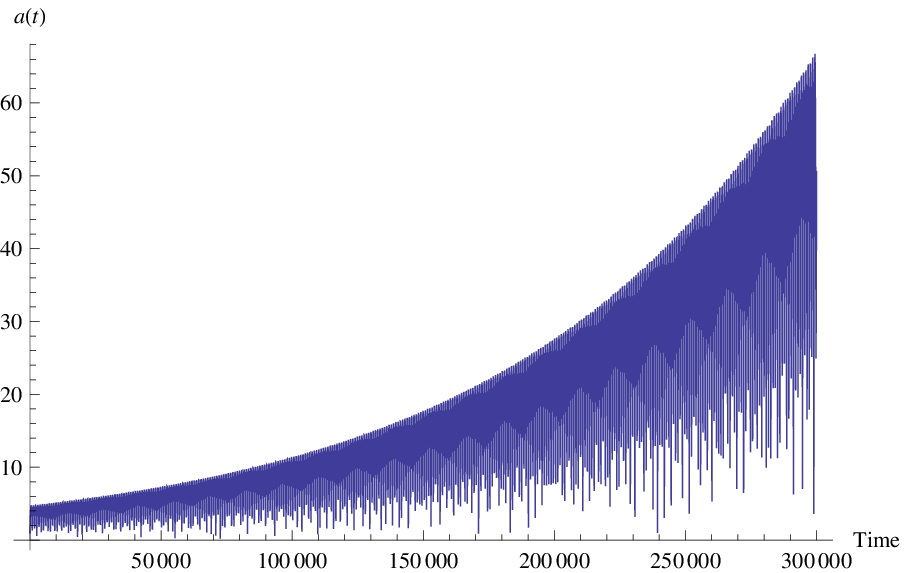}
\end{center}
\caption{{\small Left: Behavior of matter density in a single cycle. Right: The exponential growth over many cycles. Both the bounce and the turnaround scale factor grows with the same exponents, and so does their difference.
  \label{fig:growth}}}
\end{figure}
%%%%%%%%%%%%%%%%%%%%%%%%%%%

``Cyclic inflation'' (CI) can address the usual cosmological puzzles such as the ones associated with
isotropy/Mixmaster behavior, horizon, flatness and homogeneity/Black hole over-production in our universe~\cite{cyclic-inflation}, in manner very similar to the standard inflation. For instance, since
anisotropies $\propto a^{-6}$, once the cyclic-inflationary phase is ``activated'' in a small and
sufficiently smooth patch of the universe the chaotic Mixmaster behavior is avoided in
subsequent cycles because the scale factor at the consecutive bounce points also keep
growing and the universe becomes more and more isotropic. Very similar reasoning
also resolves the flatness problem.

What about the graceful exit problem? After all we have been expanding monotonically
for the last 13 billion years. In an expanding background it is known that energy densities can only decrease, and hence once the universe is in a negative energy
phase, there is no way for it to claw back up to the positive region. However,
in contracting phases the reverse is true, the energy density increases, and in~\cite{BKM-exit}
(see also~\cite{Felder,Piao:2004me}) it was demonstrated that our universe could have been ``exploring''
the negative potential regions coming from the various scalar (moduli) fields, when
at some opportune moment it was able to jump up a ``potential ladder'' to positive
potential energy regions hitherto ushering in a monotonic phase of expansion, which
is where our familiar universe finds it's place, see Fig. 2. 

Last but not the least, we need
to revisit the issue of geodesic completeness in the context of the CI scenario. If one
tracks, say, the maximum of the oscillating space time, then one finds that it has the
traditional inflationary trajectory, and the problem of past geodesic incompleteness
comes back to haunt us. Fortunately, there appears a natural resolution~\cite{B-EC} for a
closed universe: As one goes back in cycles, there comes a point when the curvature
energy density becomes more important than the vacuum energy density. (Curvature
density blue shifts as  $a^{-2}$, while the vacuum energy density remains a constant.) Once
this happens, the universe no longer turns around due to the negative vacuum energy
density, but before, when $\rho_r+\rho_k=0$. The further back we go in cycles, the universe
spends less and less time in the out-of-equilibrium phases thereby decreasing the entropy
production, and eventually, the universe asymptotes to a periodic behavior as $t\ra -\infty$~\cite{B-EC}, see Fig. 3. The space-time is, in fact, very reminiscent of the emergent
universe scenario advocated in~\cite{emergent1,emergent2,Mulryne:2005ef}, and perhaps the cosmological model should be
referred to as "emergent-cyclic-inflation".
\begin{figure}
\includegraphics[width=0.50\textwidth]{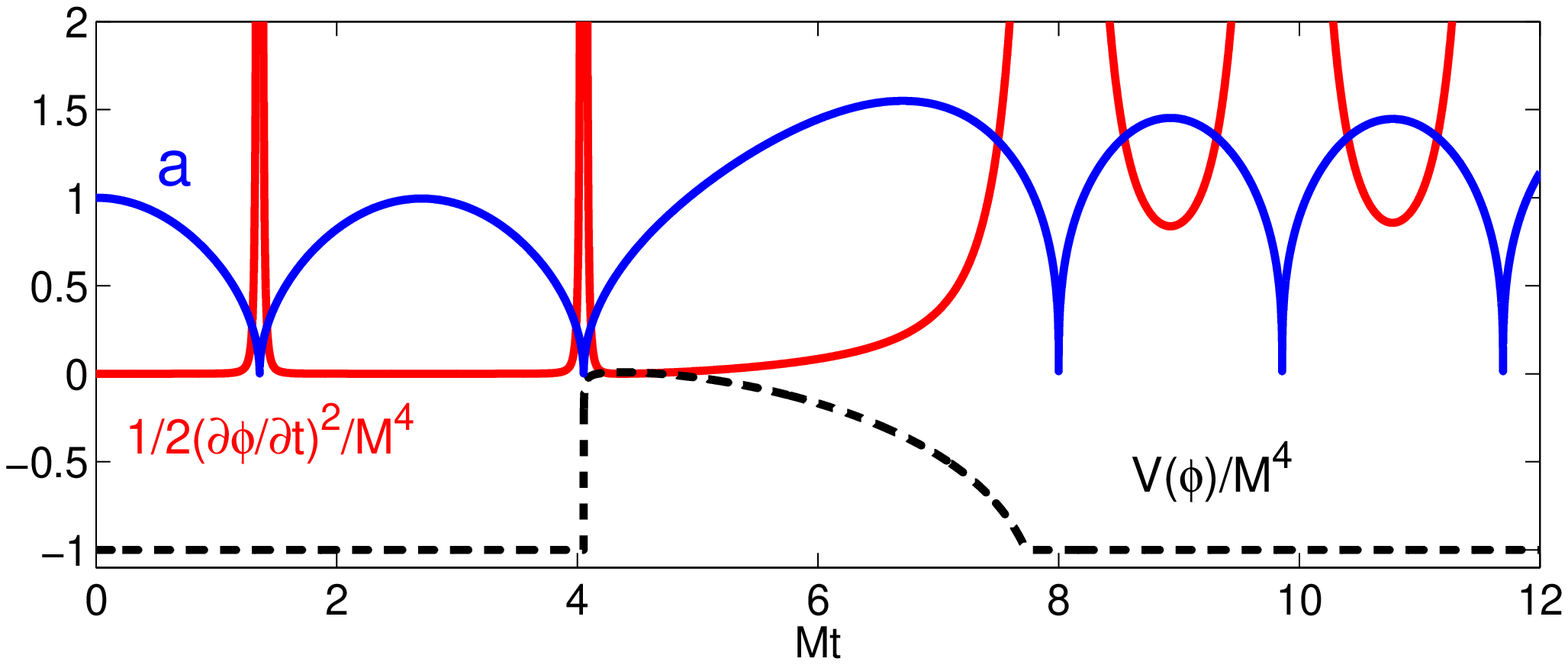}
\includegraphics[width=0.50\textwidth]{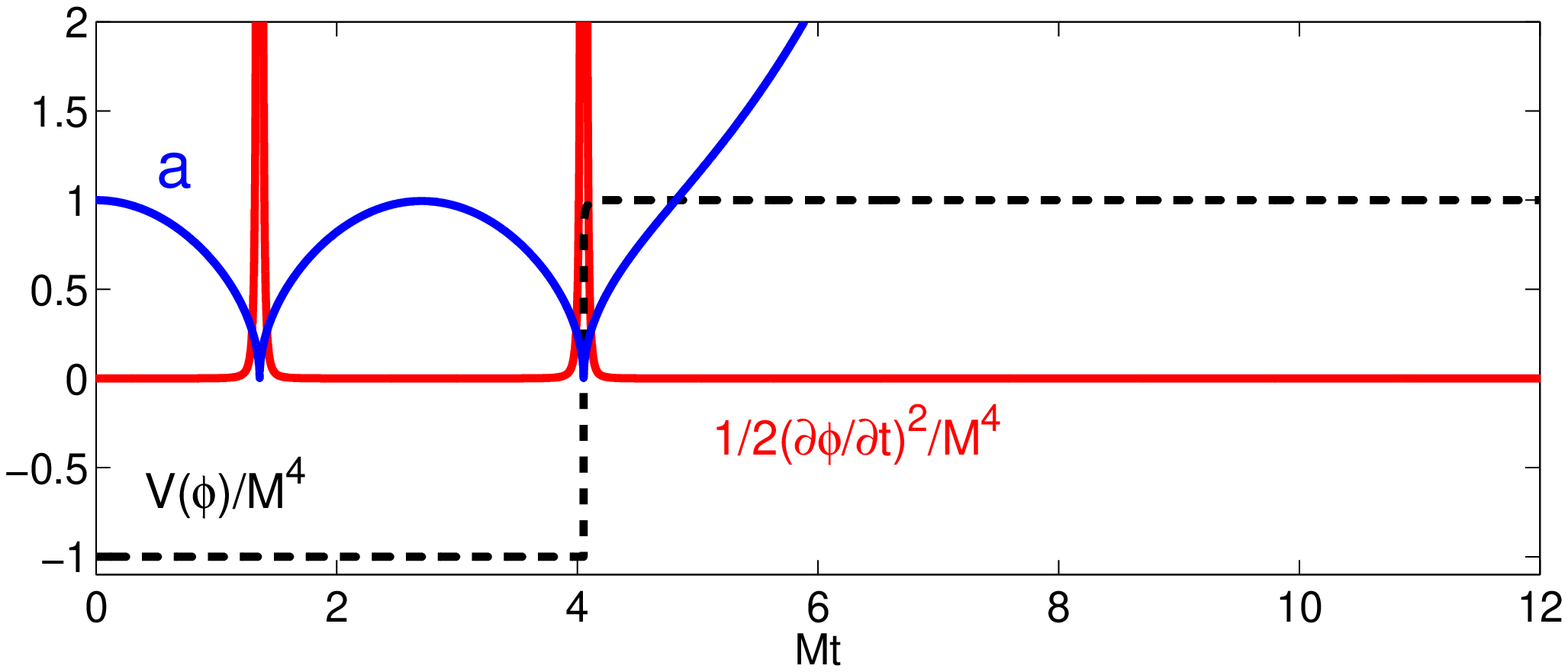}
\caption{\label{fig:evolution}
Evolution of the kinetic and potential energies. The solid blue line shows the
scale factor. The success of the graceful exit depends on the model parameters: Left:
The field reaches the upward slope in the expanding phase, and is unable to make the
transition. Right: The field is released from a different position and crosses the upward
slope completely in one bounce phase. This is the more typical evolution: in fact one has
to choose the initial conditions with some care for the field to end up in an evolution like
the left panel.}
\end{figure}
%%%%%%%%%%%%%%%%%%%%

To summarize, we have seen a viable cosmology emerges once we assume the
existence of the QG bounce, the dynamics is certainly more complex than slow-roll
inflation, but the conditions required are very reasonable, a closed universe with
negative vacuum energy (which incidentally is not particularly rare if String Theory is to be
believed), and some massless and massive excitations interacting with each other. A
good physics model though should be predictive, so for example, can one distinguish
the CI scenario from ordinary inflation? In principle, the answer is yes. It was found that the ``cyclicity'' of the inflationary phase leaves it's imprint in CMB in the form
of small logarithmic oscillations in the primordial spectrum, the fit provided in~\cite{BMS-CI}
was not definitive but encouraging. More recently, it was found that interesting
thermodynamic features, such as phase transitions, may manifest as nongaussianities
in CI models~\cite{BKM-CI}. Intriguingly, the prediction was for low $f_{NL}$ but potentially high $g_{NL}$'s,
consistent with Planck observations~\cite{Planck}. These signals do depend on the different
model parameters, and one does have to be lucky to be able to observe them in
Planck data, but perhaps it is worth investigating a bit more whether, our universe
emerged, not just from one but, from multiple bangs; before every bang there was
always another waiting right around the corner.\vs
\begin{figure}
\begin{center}
\includegraphics[height=6cm,angle=0,scale=1]{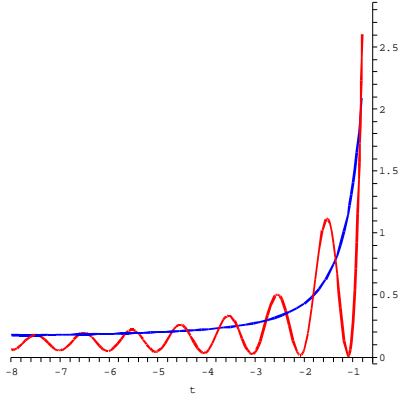}
\caption{ \label{figemergent} Qualitative plot of how the scale factor evolves as a function of the conformal time $\tau$. The cycles have the same period in $\tau$, but they grow as a function of proper time. The blue curve denotes the transition between the Hagedorn (below the curve) and the non-thermal (above the curve) phase. The evolution ends in an inflating expanding branch. }
\end{center}
\end{figure}
%%%%%%%%%%%%%%%%%%%%%%%%%%%%%
{\bf Acknowledgments:} I would like to thank  Drs. Anupam Mazumdar, Tomi Koivisto, Stephon Alexander, Arman Shaefaloo, and Mr. William Duhe for enjoyable and fruitful collaborations on the emergent-cyclic-inflation model. In many ways the present essay summarizes some of the important findings and ideas of these collaborations. I would especially like to thank Anupam for his helpful comments during the preparation of the article. This work was supported by the LEQSF(2011-13)-RD-A21 grant from the Louisiana Board of Regents.

%%%%%%%%%%%%%%%%%%%%%%%%%%%%
\bibliography{cyclicrefs}

\begin{thebibliography}{10}

\bibitem{short-distance-geraci}
A.~A. Geraci, S.~J. Smullin, D.~M. Weld, J.~Chiaverini, and A.~Kapitulnik,
  ``{Improved constraints on non-Newtonian forces at 10 microns},'' {\em
  Phys.Rev.}, vol.~D78, p.~022002, 2008.

\bibitem{short-distance-weld}
D.~M. Weld, J.~Xia, B.~Cabrera, and A.~Kapitulnik, ``{A New Apparatus for
  Detecting Micron-Scale Deviations from Newtonian Gravity},'' {\em Phys.Rev.},
  vol.~D77, p.~062006, 2008.

\bibitem{short-distance-newman}
R.~Newman, E.~Berg, and P.~Boynton, ``{Tests of the gravitational inverse
  square law at short ranges},'' {\em Space Sci.Rev.}, vol.~148, pp.~175--190,
  2009.

\bibitem{WMAP}
E.~Komatsu {\em et~al.}, ``{Seven-Year Wilkinson Microwave Anisotropy Probe
  (WMAP) Observations: Cosmological Interpretation},'' {\em Astrophys. J.
  Suppl.}, vol.~192, p.~18, 2011.

\bibitem{Planck}
P.~Ade {\em et~al.}, ``{Planck 2013 results. XVI. Cosmological parameters},''
  2013.

\bibitem{Guth}
A.~Borde, A.~H. Guth, and A.~Vilenkin, ``{Inflationary space-times are
  incompletein past directions},'' {\em Phys.Rev.Lett.}, vol.~90, p.~151301,
  2003.

\bibitem{Borde}
A.~Borde and A.~Vilenkin, ``{Eternal inflation and the initial singularity},''
  {\em Phys.Rev.Lett.}, vol.~72, pp.~3305--3309, 1994.

\bibitem{Linde}
A.~D. Linde, D.~A. Linde, and A.~Mezhlumian, ``{From the Big Bang theory to the
  theory of a stationary Universe},'' {\em Phys.Rev.}, vol.~D49,
  pp.~1783--1826, 1994.

\bibitem{BMS}
T.~Biswas, A.~Mazumdar, and W.~Siegel, ``{Bouncing universes in string-inspired
  gravity},'' {\em JCAP}, vol.~0603, p.~009, 2006.

\bibitem{Ashtekar}
A.~Ashtekar, T.~Pawlowski, and P.~Singh, ``{Quantum Nature of the Big Bang:
  Improved dynamics},'' {\em Phys.Rev.}, vol.~D74, p.~084003, 2006.

\bibitem{loop}
M.~Bojowald, ``{Loop quantum cosmology},'' {\em Living Rev.Rel.}, vol.~8,
  p.~11, 2005.

\bibitem{Shtanov}
Y.~Shtanov and V.~Sahni, ``{Bouncing brane worlds},'' {\em Phys.Lett.},
  vol.~B557, pp.~1--6, 2003.

\bibitem{Freese}
K.~Freese, M.~G. Brown, and W.~H. Kinney, ``{The Phantom Bounce: A New Proposal
  for an Oscillating Cosmology},'' 2008.

\bibitem{Baum}
L.~Baum and P.~H. Frampton, ``{Turnaround in cyclic cosmology},'' {\em
  Phys.Rev.Lett.}, vol.~98, p.~071301, 2007.
\newblock Altered and improved model astro-ph/0608138 with new title.

\bibitem{palatini-bounce}
T.~S. Koivisto, ``{Bouncing Palatini cosmologies and their perturbations},''
  {\em Phys.Rev.}, vol.~D82, p.~044022, 2010.

\bibitem{ekpyrotic}
J.~Khoury, B.~A. Ovrut, P.~J. Steinhardt, and N.~Turok, ``{The Ekpyrotic
  universe: Colliding branes and the origin of the hot big bang},'' {\em
  Phys.Rev.}, vol.~D64, p.~123522, 2001.

\bibitem{matter-bounce}
R.~H. Brandenberger, ``{The Matter Bounce Alternative to Inflationary
  Cosmology},'' 2012.

\bibitem{BBMS}
T.~Biswas, R.~Brandenberger, A.~Mazumdar, and W.~Siegel, ``{Non-perturbative
  Gravity, Hagedorn Bounce \& CMB},'' {\em JCAP}, vol.~0712, p.~011, 2007.

\bibitem{supernova}
S.~Perlmutter {\em et~al.}, ``{Discovery of a Supernova Explosion at Half the
  Age of the universe and its Cosmological Implications},'' {\em Nature},
  vol.~391, pp.~51--54, 1998.

\bibitem{tolman1}
R.C.Tolman, ``{Relativity,Thermodynamics and Cosmology},'' {\em Oxford U.Press,
  Clarendon Press}, 1934.

\bibitem{tolman2}
R.C.Tolman, ``{On the Problem of the Entropy of the Universe as a Whole},''
  {\em Phys.Rev.}, vol.~37, p.~1639, 1931.

\bibitem{cyclic-inflation}
T.~Biswas and A.~Mazumdar, ``{Inflation with a negative cosmological
  constant},'' {\em Phys.Rev.}, vol.~D80, p.~023519, 2009.

\bibitem{BD}
W.~Duhe and T.~Biswas, ``{Emergent Cyclic Inflation, a Numerical
  Investigation},'' 2013.

\bibitem{BKM-exit}
T.~Biswas, T.~Koivisto, and A.~Mazumdar, ``{Could our Universe have begun with
  Negative Lambda?},'' 2011.

\bibitem{Felder}
G.~N. Felder, A.~V. Frolov, L.~Kofman, and A.~D. Linde, ``{Cosmology with
  negative potentials},'' {\em Phys.Rev.}, vol.~D66, p.~023507, 2002.

\bibitem{Piao:2004me}
Y.-S. Piao, ``{AdS minima and anthropic cycles of universe},'' {\em Phys.
  Rev.}, vol.~D70, p.~101302, 2004.

\bibitem{B-EC}
T.~Biswas, ``{The Hagedorn Soup and an Emergent Cyclic Universe},'' 2008.

\bibitem{emergent1}
G.~F. Ellis and R.~Maartens, ``{The emergent universe: Inflationary cosmology
  with no singularity},'' {\em Class.Quant.Grav.}, vol.~21, pp.~223--232, 2004.

\bibitem{emergent2}
G.~F. Ellis, J.~Murugan, and C.~G. Tsagas, ``{The Emergent universe: An
  Explicit construction},'' {\em Class.Quant.Grav.}, vol.~21, pp.~233--250,
  2004.

\bibitem{Mulryne:2005ef}
D.~J. Mulryne, R.~Tavakol, J.~E. Lidsey, and G.~F. Ellis, ``{An Emergent
  Universe from a loop},'' {\em Phys.Rev.}, vol.~D71, p.~123512, 2005.

\bibitem{BMS-CI}
T.~Biswas, A.~Mazumdar, and A.~Shafieloo, ``{Wiggles in the cosmic microwave
  background radiation: echoes from non-singular cyclic-inflation},'' {\em
  Phys.Rev.}, vol.~D82, p.~123517, 2010.

\bibitem{BKM-CI}
T.~Biswas, T.~Koivisto, and A.~Mazumdar, ``{Phase transitions during
  Cyclic-Inflation and Non-gaussianity},'' 2013.

\end{thebibliography}
\bibliographystyle{ieeetr}
\end{document}